\documentclass[11pt,a4paper]{article}
\usepackage{amsmath, amssymb}
\usepackage{latexsym}
\usepackage{graphicx, epsfig}
\usepackage{color}
\usepackage{algorithm,algcompatible}

\usepackage{caption}
\usepackage{subcaption}

\newcommand{\bef}{\begin{figure}}
\newcommand{\enf}{\end{figure}}

\numberwithin{equation}{section}

\newtheorem{thm}{Theorem}[section]

\newcommand{\p}{\partial}

\newcommand{\beq}{\begin{equation}}
\newcommand{\eeq}{\end{equation}}

\title{Asymptotic analysis of the Narrow Escape Problem in general shaped domain with several absorbing necks}

\author{ Xiaofei Li\thanks{College of science, Zhejiang University of Technology, Hangzhou, 310023, P. R. China (xiaofeilee@hotmail.com, shengqilin@zjut.edu.cn). Corresponding author.}\and
Shengqi Lin\thanks{College of science, Zhejiang University of Technology, Hangzhou, 310023, P. R. China (shengqilin@zjut.edu.cn).}
}

\begin{document}

\maketitle

\begin{abstract}

This paper considers the two-dimensional narrow escape problem in a domain which is composed of a relatively big head and several absorbing narrow necks. The narrow escape problem is to compute the mean first passage time(MFPT) of a Brownian particle traveling from inside the head to the end of the necks. The original model for MFPT is to solve a mixed Dirichlet-Neumann boundary value problem for the Poisson equation in the composite domain, and is computationally challenging.  In this paper, we compute the MFPT by solving an equivalent Neumann-Robin type boundary value problem. By solving the new model, we obtain the three-term asymptotic expansion of the MFPT. We also conduct numerical experiments to show the accuracy of the high order expansion. As far as we know, this is the first result on high order asymptotic solution for NEP in a general shaped domain with several absorbing neck windows. This work is motivated by \cite{Li}, where the Neumann-Robin model was proposed to solve the NEP in a domain with a single absorbing neck. 

\end{abstract}

\noindent{\footnotesize {\bf Key words.} Narrow escape problem; Mean first passage time; Asymptotic analysis; Several absorbing necks; Neumann-Robin model}

\section{Introduction}

When a Brownian particle is confined to a simply connected bounded domain $\Omega$ with a small absorbing window and the other part of the boundary is reflecting, it attempts to escape from the domain through this small absorbing window. The time that the Brownian particle takes from an initial position $x\in \Omega$ to escape via the absorbing window is called the mean first passage time(MFPT). The narrow escape problem(NEP) is to calculate the MFPT of the confined particle. Let $\Omega$ be a bounded simply connected domain in $\mathbb{R}^2$ with $C^2$-smooth  boundary $\p\Omega$. The boundary $\p\Omega$ is composed of two disjoint parts, the absorbing part $\p\Omega_a$ and the reflecting part $\p\Omega_r$, such that $\p\Omega =\p\Omega_a\cup \p\Omega_r$. We assume that $|\p \Omega_a| = O(\epsilon)$ for $\epsilon\ll 1$, while $\p\Omega$ is of order $1$. Suppose that a Brownian particle is confined at the position $x\in \Omega$. The MFPT $u(x)$, which depends on the starting position $x\in\Omega$, satisfies the following equation:
\beq\label{model}
\begin{cases}
\Delta u(x)=-\frac{1}{D}\quad &\mbox{in}~\Omega,\\[1ex]
\dfrac{\partial u}{\partial \nu}(x)=0&\mbox{on}~\partial \Omega_{r},\\[1.5ex] 
u(x) = 0 &\mbox{on\ } \partial \Omega _{a},
\end{cases}
\eeq
where $\nu$ is the outer unit normal to $\p\Omega$ and $D$ is the diffusion coefficient. In this paper, we consider the simplest form of pure diffusion with a unit diffusion coefficient $D=1$. The asymptotic analysis for NEP arises in deriving the asymptotic expansion of $u$ as $\epsilon\rightarrow 0$, from which one can estimate the MFPT of the confined particle. The above Dirichlet-Neumann type boundary value problem \eqref{model} can be derived by considering the probability density function of the Brownian particle at location $x$ at time $t$ through Fokker-Planck equation. More details on the derivation can be found in \cite{AKKL}.

The NEP has attracted significant attention from the point view of mathematical and numerical concern due to its application in molecular biology and biophysics. There have been significant works in deriving the leading order term and higher order terms of the asymptotic expansion of MFPT in a regular bounded domain in two and three dimensions, see, for example,  \cite{AKKL,michael,HSbook,cluster,laws,planar,PWPK,leakage}. Most existing results are focusing on the derivation of the MFPT in a domain with a single small absorbing gate. Whereas, a few works have been conducted for a domain with multiple small absorbing windows. The leading order term of the asymptotic formula of MFPT in a smooth bounded domain with multiple small absorbing windows has been studied in \cite{agkls,numer,DWC,review,LH,LBW}.

In this paper, we rigorously derive the high leading order asymptotic expansion of the MFPT in a domain which is composed of a relatively big head and several narrow absorbing necks. In fact, the MFPT in such domain is closely related to the diffusion of particles in a cellular network since the cellular network are mostly composed of narrow necks connecting relatively larger head compartments. The shape and the distribution of the head compartments as well as the distribution of the tubules are involved in processes including active or diffusion transport of proteins, calcium signaling and so on \cite{networks}. How changes in shape of the compartment or the ratio of head to necks occur in response to specific cellular signals is important. However, how these structures regulate molecular trafficking and diffusion is unclear \cite{DM}. In \cite{Li}, we considered a new model of Robin-Neumann type boundary value problem to compute the MFPT in a domain composed of a relatively big head and a single neck gate. Motivated by \cite{Li} and its possible implications on diffusion of particles in cellular networks, we consider the NEP problem in a domain which is composed of a relatively big head and several thin absorbing necks from mathematical point of view in this paper.

\begin{figure}[!ht]
	\centering
	\includegraphics[width=60mm]{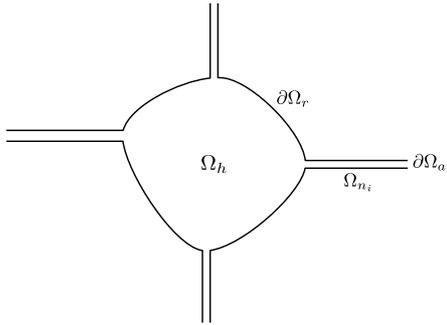}
	\caption{ A head domain connected with $N$ narrow necks, where $N=4$.}
	\label{fig:Network domain_partial}
\end{figure}

Let $\Omega_h$ be a simply connected bounded domain with a $C^2$ boundary. Let $\Omega_{n_i}$ be rectangular neck with length $L_i$ and width $2\epsilon_i$, where $\epsilon_i \ll 1$, for $i = 1,\dots,N$. Each $\Omega_{n_i}$ is connected to $\Omega_h$ with connecting segment $\Gamma_{\epsilon_i}$. We assume that $\Gamma_{\epsilon_i} \cap \Gamma_{\epsilon_j}=\emptyset$, $i\neq j$, and moreover $dist(\Gamma_{\epsilon_i},\Gamma_{\epsilon_j})\geq c$, for some $c=O(1)$, in other words, they are well-separated. Let $\Omega = \Omega_h \cup \Omega_{n_1} \cup \dots \cup \Omega_{n_N}$. The boundary of $\Omega$ is divided into two parts, one is the reflecting boundary $\p\Omega_r$, and the other absorbing boundary $\p\Omega_{a}$, which consists of $N$ parts $\p\Omega_{a_i}$, $i = 1,,\dots,N$, where $\p\Omega_{a_i}$ is the end of each neck. We assume that the size of each $\p\Omega_{a_i}$ is much smaller than that of the whole boundary $\p\Omega$. As an example,  the geometric description for four necks is shown in Figure~\ref{fig:Network domain_partial}. The NEP is to calculate the MFPT $u$ which is the unique solution to the following Dirichlet-Neumann model:
\beq\label{original_u}
\begin{cases}
\Delta u(x)=-1\quad &\mbox{in}~\Omega,\\[1ex]
\dfrac{\partial u}{\partial \nu}(x)=0&\mbox{on}~\partial \Omega_{r},\\[1.5ex] 
u(x) = 0 &\mbox{on\ } \partial \Omega _{a_i},
\end{cases}
\eeq
for all $i = 1,\dots,N$, where $\nu$ is the outward unit normal to $\p \Omega$. In this paper, we derive the high order asymptotic expansion of $u$ as $\max(\epsilon_1,\dots,\epsilon_N)\rightarrow 0$, from which one can estimate the escape time of the Brownian particle through the absorbing neck gate with high accuracy. In fact, three leading order term asymptotic solution to (\ref{original_u}) is derived by means of solving an equivalent Neumann-Robin type boundary value problem, which is proposed in \cite{Li} to calculate the MFPT in a domain with a single absorbing neck window. 

The main idea of the Neumann-Robin model is as follows:  since the neck is relatively thin, it can be approximated as one dimensional, then the solution $u$ only varies along the neck direction. By dropping the neck and using the continuity of $u$ and its derivatives, it is technically equivalent as considering a Robin boundary condition. Following the same spirit, in this paper, we drop all the necks and instead consider Robin boundary condition at each connecting segment $\Gamma_{\epsilon_i}$, $i = 1,\dots,N$. Hence, we can reformulate the original problem \eqref{original_u} as the following Neumann-Robin model:
\beq \label{NR_eq}
\begin{cases}
\Delta u(x)=-1\quad &\mbox{in}~\Omega_h,\\[1ex]
\dfrac{\partial u}{\partial \nu}(x)=0&\mbox{on}~\partial \Omega_{r},\\[1em]
\dfrac{\partial u}{\partial \nu}(x)+\alpha_i u(x)=\beta_i &\mbox{on}~\Gamma_{\epsilon_i}.
\end{cases}
\eeq
Here $\alpha_i$ and $\beta_i$ are constants which are determined in Section 3. We assume that each $\epsilon_i$ is sufficiently small such that $\alpha_i\epsilon_i \ll 1$. 

By considering the Neumann-Robin model \eqref{NR_eq}, we transform the problem from a singular domain $\Omega$ into a smooth domain $\Omega_h$ by dropping all the necks and assigning Robin condition on each connecting segment. In this way, we shall use layer potential techniques to derive the high order asymptotic solution $u$ to (\ref{NR_eq}), and obtain the following main theorem of this paper.

\begin{thm}\label{main_thm}
Let $x\in \Omega_h$ and be away from $\Gamma_{\epsilon_i}$, $i = 1,\dots,N$. The MFPT for a Brownian particle escaping from the initial position $x\in \Omega_h$ to the neck gate, i.e., the solution $u$ to \eqref{original_u} is
$$
u(x)=\frac{\left| \Omega _h \right|}{2\sum_{i=1}^N{\epsilon _i/L_i}}+\frac{\left| \Omega _h \right|}{\pi}\left( \sum_{i=1}^{N-1}{\sum_{j=i+1}^N{T_{ij}\ln \epsilon _i\epsilon _j}}-\sum_{i=1}^N{F_i\ln \epsilon _i} \right)+O\left( 1 \right),
$$
where the constants $T_{ij}$, $F_i$ are given by
$$
T_{ij}=\frac{\epsilon _i\epsilon _j}{L_iL_j}\Big/\left( \sum_{k=1}^N{\frac{\epsilon _k}{L_k}} \right)^2,~ F_i=\frac{\epsilon _i}{L_i}\Big/\sum_{k=1}^N{\frac{\epsilon _k}{L_k}}.
$$
\end{thm}
Moreover, if $\Omega_h$ is a unit disk centered at $(0,0)$, the necks have the same length $L$ and the same width $2\epsilon$, then the MFPT $u$ has the following explicit formula up to order $O\left( \epsilon \ln ^2\epsilon \right)$:
\beq\label{un}
\begin{aligned}
u\left( x \right) =&\frac{\left| \Omega _h \right|L}{2N}\frac{1}{\epsilon}-\frac{\left| \Omega _h \right|}{\pi N}\ln \epsilon -\frac{\left| \Omega _h \right|}{2\pi N}\left( 2\ln 2-3 \right)+\frac{L^{2}}{2}-\frac{2\left| \Omega _h \right|}{\pi N^2}\sum_{i=1}^{N-1}\sum_{j=i+1}^N \ln \left| s_i-s_j \right|\\
&+\frac{1}{4}\left( 1-\left| x \right|^2 \right) +\frac{\left| \Omega _h \right|}{\pi N}\sum_{i=1}^N{\ln \left| x-s_i  \right|}+O\left( \epsilon \ln ^2\epsilon \right),
\end{aligned}
\eeq
where $s_i$ is the center point on $\Gamma_{\epsilon_i}$. The first leading term of \eqref{un} is proportional to the length $L$ and of order $O(1/\epsilon)$, which is different from the well-known leading term $O(\ln \epsilon)$ for two dimensional NEP without necks. Due to the existence of narrow necks, it takes longer time for the particle to escape, which is natural and also interesting. Moreover, the three leading terms are explicitly given in formula \eqref{un}. The formula \eqref{un} can approximate the MFPT with high accuracy, which can be seen from the numerical results in Section \ref{Numerical}.

It is also worth mentioning that, let $N=1$, i.e., there is only a single neck gate, then the above formula becomes 
\beq\label{u1}
\begin{aligned}
u\left( x \right) &=\frac{\left| \Omega _h \right|L}{2}\frac{1}{\epsilon}-\frac{\left| \Omega _h \right|}{\pi }\ln \epsilon -\frac{\left| \Omega _h \right|}{2\pi }\left( 2\ln 2-3 \right)+\frac{L^{2}}{2}\\
&+\frac{1}{4}\left( 1-\left| x \right|^2 \right) +\frac{\left| \Omega _h \right|}{\pi }{\ln \left| x-s_1  \right|}+O\left( \epsilon  \right),
\end{aligned}
\eeq
which is exactly the same as the result in \cite{Li}, where the MFPT was derived for a single neck exit. 

By comparing \eqref{un} and \eqref{u1}, one can observe that the first and second leading order terms of \eqref{un}, where there exist $N$ necks, is $1/N$ of that of \eqref{u1}, where there exists only a single neck, which is quite natural. But, the third leading order term $O(1)$ does not satisfy the $1/N$ relation between \eqref{un} and \eqref{u1}. The reason is that the $O(1)$ term depends not only on the location of the starting point, the length of the neck, but also on the location of the narrow necks as well as the nonlinear interaction between them. We will show how the $O(1)$ term contributes to the accuracy of the MFPT through numerical results in Section \ref{Numerical}. As far as we know, this paper gives the first result on the three-term asymptotic expansion of the solution to NEP in a domain with several thin neck windows.

This paper is organized as follows. In section 2, we review the Neumann function for the Laplacian in $\mathbb{R}^2$. In section 3, we derive an equivalent Neumann-Robin model for the associated NEP in a domain with several thin necks. In section 4, we rigorously derive the high order asymptotic solution to the Neumann-Robin model by using layer potential techniques. Numerical experiments are provided in Section \ref{Numerical} to confirm the theoretical results. This paper ends with a short conclusion.

\section{Neumann function in $\mathbb{R}^2$}

In this section, we review on the structure of Neumann function for a regular domain in $\mathbb{R}^2$ for further use, refer to \cite{HSbook}.

Let $\Omega$ be a bounded domain in $\mathbb{R}^2$ with $C^2$ smooth boundary $\partial\Omega$, and let $G(x,z)$ be the Neumann function for $-\Delta$ in $\Omega$ with a given $z\in\Omega$. That is, $G(x,z)$ is the solution to the boundary value problem
\begin{equation*}
  \begin{cases}
        \Delta_{x} G(x,z)=-\delta_{z}, &x\in \Omega,\\[1ex]
      \displaystyle\frac{\partial G}{\partial \nu_{x}}=-\frac{1}{|\partial \Omega|},  & x\in \partial\Omega,\\[1em]
      \displaystyle\int_{\partial \Omega}G(x,z)d\sigma(x)=0,&
  \end{cases}
\end{equation*}
where $\nu$ is the outer unit normal to the boundary $\partial\Omega$.

If $z\in \Omega$, then $G(x,z)$ can be written in the form
\begin{eqnarray*}
G(x,z)=-\frac{1}{2\pi} \ln |x-z| +R_{\Omega}(x,z),\quad x\in\Omega,
 \end{eqnarray*}
where $R_{\Omega}(x,z)$ is the regular part which belongs to $H^{3/2}(\Omega)$, the standard $L^2$ Sobolev space of order $3/2$, which solves the boundary value problem
\begin{equation*}
  \begin{cases}
-\Delta_{x} R_{\partial\Omega}(x,z)=0,&x\in\Omega,\\[1ex]
\displaystyle\frac{\partial R_{\Omega}}{\partial \nu_{x}}\Big|_{x\in \partial \Omega}=-\frac{1}{|\partial \Omega|}+\frac{1}{2\pi}\frac{\langle x-z,\nu_{x}\rangle}{|x-z|^{2}},&x\in \partial \Omega,
\end{cases}
\end{equation*}
where $\langle\cdot,\cdot\rangle$ denotes the inner product in $\mathbb{R}^2$.

If $z\in \partial \Omega$, then Neumann function on the boundary is denoted by $G_{\partial \Omega}$ and can be written as
\begin{align}\label{N}
G_{\partial \Omega}(x,z)=-\frac{1}{\pi} \ln |x-z| + R_{\partial\Omega}(x,z),\quad x \in \Omega,~z \in \partial \Omega,
\end{align}
where $R_{\partial \Omega}(x,z)$ has weaker singularity than $\ln |x-z|$ and solves the boundary value problem
\begin{equation*}
  \begin{cases}
        \Delta_{x}R_{\partial \Omega}(x,z)=0,& x\in \Omega,\\[1ex]
       \displaystyle\frac{\partial R_{\partial\Omega}}{\partial \nu_{x}}\Big|_{x\in \partial \Omega}=-\frac{1}{|\partial \Omega|}+\frac{1}{\pi}\frac{\langle x-z,\nu_{x}\rangle}{|x-z|^{2}},&x\in \partial \Omega, ~z\in \partial \Omega.
       \end{cases}
\end{equation*}

In particular, if $\Omega$ is the unit disk, then
$$\frac{\langle x-z,\nu_{x}\rangle}{|x-z|^2}=\frac{\langle x-z,x \rangle}{|x|^2+|z|^2-2x\cdot z}=\frac{1-x\cdot z}{2-2x\cdot z}=\frac{1}{2}$$
for $x\in \partial\Omega$. Therefore, $\frac{\partial R_{\partial\Omega}}{\partial \nu_{x}}\Big|_{x\in \partial \Omega}=0$, and hence $R_{\partial \Omega}(x,z)=$constant. Since $\int_{\partial \Omega}G(x,z)d\sigma(x)=0$, we have $R_{\partial \Omega}(x,z)=0$ for all $x\in\Omega$ and $z\in\partial\Omega$, and hence
\begin{eqnarray*}
G_{\partial \Omega}(x,z)=-\frac{1}{\pi}\ln|x-z|,~x\in\Omega,~z\in\partial\Omega.
\end{eqnarray*}
We also have
\begin{eqnarray*}
\int_{\Omega} G(x,z)dz=\frac{1}{4}(1-|x|^2).
\end{eqnarray*}

For later use, we introduce the integral operator $P:L^2[-1,1]\rightarrow L^2[-1,1]$, defined by
$$P[\phi](x)=\int_{-1}^{1}\ln|x-y|\phi(y)dy.$$
The operator $P$ is bounded (see Lemma 2.1 in \cite{AKKL}). The following quantity (obtained in \cite{Li}) is useful for later calculation
\beq\label{L1}
\int_{-1}^1{P\left[ 1 \right]}\left( t \right) dt=4\ln 2-6.
\eeq

\section{Derivation of the equivalent Neumann-Robin model}

In this section, we derive an equivalent Neumann-Robin type boundary value problem to \eqref{original_u} to solve the NEP in a composite domain which comprises a relatively large head and several narrow rectangular absorbing necks. 

Let $\Omega_h$ be a simply connected bounded domain with a $C^2$ boundary. Let $\Omega_{n_i}$ be rectangular neck with length $L_i$ and width $2\epsilon_i$, where $\epsilon_i \ll 1$, for $i = 1,\dots,N$. The necks are connected to the head domain $\Omega_h$ and they are well-separated. The connecting part between $\Omega_h$ and $\Omega_{n_i}$ is a small line segment which is denoted by $\Gamma_{\epsilon_i}$. Let $\p \Omega_{a_i}$ be the exit, which is the end of the neck $\Omega_{n_i}$. Denote $\Omega: = \Omega_h \cup \Omega_{n_1} \cup \dots \cup \Omega_{n_N}$.

Consider the neck domain $\Omega_{n_i}$. Take the center point of $\Gamma_{\epsilon_i}$ as the point of origin $(0,0)$, then the neck direction is along the $x$- axis. Following \eqref{original_u}, the MFPT $u$ in domain $\Omega_{n_i}$ should satisfy the boundary value problem:
\begin{equation*}
\begin{cases}
\Delta u(x)=-1\quad &\mbox{in}~\Omega_{n_i},\\[1ex]
\dfrac{\partial u}{\partial \nu}(x)=0&\mbox{on}~\partial \Omega_{r},\\[1.5ex] 
u(x) = 0 &\mbox{on\ } \partial \Omega _{a_i}, \\[0.5ex] 
u(x) = u(x,y) &\mbox{on\ } \Gamma_{\epsilon_i},
\end{cases}
\end{equation*}
where $x$ is the $x$- coordinate and $y$ is the $y$- coordinate in the two-dimensional domain. 

As it is noted in paper \cite{Li}, since the width of the neck is small, the neck domain $\Omega_{n_i}$ can be approximated as one-dimensional. The solution $u$ only varies along the $x$- coordinate, and thus solves the following ordinary differential equation:
\begin{equation*}
\begin{aligned}
&\frac{d^2 u(x,y)}{dx^2} =-1 \quad~\mbox{for}~ x\in [0,L_i], \\
&u(x,y) = 0 \quad\quad~\mbox{for}~ x = 0,\\
&u(x,y) = C_i  \quad\quad~\mbox{for}~x = L_i,
 \end{aligned}
 \end{equation*}
where $C_i$ is constant. By separation of variables, we can solve the above ODE and obtain 
\beq\label{ux}
u(x,y) =  -\frac{1}{2}(L_i -x)^2 + \left(\frac{C_i}{L_i} + \frac{L_i}{2}\right)(L_i-x),
\eeq
where $x\in [0,L_i]$, $y\in [-\epsilon_i,\epsilon_i]$.

Since $u$ and the derivative of $u$ are both continuous across the connecting segment $\Gamma_{\epsilon_i}$, they should satisfy a Robin type condition 
\beq\label{R}
\frac{\p u}{\p \nu} + \alpha_i u = \beta_i  \quad \mbox{on} ~ \Gamma_{\epsilon_i},
\eeq
where $\alpha_i$, $\beta_i$ are to be determined. In fact, substituting \eqref{ux} into \eqref{R}, one can see that 
\beq\label{ab}
\alpha_i=\frac{1}{L_i},~\beta_i = \frac{L_i}{2}.
\eeq

Therefore, by \eqref{R} and \eqref{ab}, the MFPT $u$ for a confined particle from the initial position $x$ in the head domain $\Omega_h$ to escape from the neck exits, is to solve the following equivalent  Neumann-Robin boundary value problem:

\beq \label{NR}
\begin{cases}
\Delta u(x)=-1\quad &\mbox{in}~\Omega_h,\\[1ex]
\dfrac{\partial u}{\partial \nu}(x)=0&\mbox{on}~\partial \Omega_{r},\\[1em]
\dfrac{\p u}{\p \nu} + \dfrac{1}{L_i} u = \dfrac{L_i}{2} &\mbox{on}~\Gamma_{\epsilon_i}.
\end{cases}
\eeq
Note that, by considering the Neumann-Robin model, we transform a singular domain $\Omega_h\cup \Omega_{n_1}\cup \dots \cup \Omega_{n_N}$ into a smooth domain $\Omega_h$ by dropping all the necks and assigning Robin condition on each connecting segment. In this way, we can derive high order asymptotic expansion of the MFPT by using layer potential techniques in the smooth domain.

\section{Derivation of the three-term asymptotic expansion of MFPT}

In this section, we prove the main Theorem \ref{main_thm} in a smooth domain $\Omega_h$, where $\p \Omega_h \in C^2(\mathbb{R}^2)$. The high order asymptotic expansion of MFPT $u$ is rigorously derived by solving the equivalent Neumann-Robin model \eqref{NR} using layer potential techniques.

Proof.  Let 
$$g(x) = \int_{\Omega_h} G(x,z)dz, ~x\in \Omega_h,$$
then $g$ solves the following boundary value problem
\beq\label{g}
\begin{cases}
\Delta g_(x)=-1\quad &\mbox{in}~\Omega_h,\\[1ex]
\dfrac{\partial g}{\partial \nu}(x)=-\dfrac{\left| \Omega _h \right|}{\left| \partial \Omega _h \right|} &\mbox{on}~\partial \Omega_{h},\\[1em] 
\int_{\partial \Omega _h}{g}d\sigma =0.
\end{cases}
\eeq

Applying Green's second formula, and using \eqref{NR} and \eqref{g}, we obtain 
\beq\label{u_multi}
u\left( x \right) =g\left( x \right) +\sum_{i=1}^N {\int_{\Gamma _{\epsilon_i}}G_{\partial \Omega}\left( x,y \right)  {\frac{\partial u\left( y \right)}{\partial \nu}d\sigma \left( y \right)}}+C_{\epsilon},
\eeq
where the constant $C_{\epsilon}$ is given by
$$C_{\epsilon}=\frac{1}{\left| \partial \Omega _h \right|}\int_{\partial \Omega _h}{u\left( y \right) d\sigma \left( y \right)}.$$ 
Here, $\frac{\p u}{\p \nu}$ on $\Gamma_{\epsilon_i}$, $i = 1,2,\dots,N$, and $C_\epsilon$ are to be determined.

By \eqref{N}, and the third and fourth Robin boundary conditions in \eqref{NR},  one can see that \eqref{u_multi} can be written as
\beq\label{u_original}
\begin{aligned}
\frac{L_{j}^{2}}{2}-L_j\frac{\partial u\left( x \right)}{\partial \nu}=&g\left( x \right) -\frac{1}{\pi}\sum_{i=1}^N \int_{\Gamma_{\epsilon_i}}{\ln \left| x-y \right|\frac{\partial u\left( y \right)}{\partial \nu}d\sigma \left( y \right)} \\
&+\sum_{i=1}^N \int_{\Gamma_{\epsilon_i}}{R_{\partial\Omega _h}\left( x,y \right) \frac{\partial u\left( y \right)}{\partial \nu}d\sigma \left( y \right)} + C_\epsilon,
\end{aligned}
\eeq
for $x\in \Gamma_{\epsilon_j}$, $j=1,2,\dots,N$, respectively.

Let $s_i$ be the center point on $\Gamma_{\epsilon_i}$, $i = 1,2,\dots, N$, and let $x_i(t)$: $[-\epsilon_i + s_i, \epsilon_i + s_i] \rightarrow \mathbb{R}^2$ be the arc-length parametrization of $\Gamma_{\epsilon_i}$, i.e., $|x_i'(t)| = 1$ for all $t \in [-\epsilon_i + s_i, \epsilon_i + s_i]$. Then 
$$\Gamma_{\epsilon_i} = \{x_i(t)| t\in [-\epsilon_i + s_i, \epsilon_i + s_i]\}.$$
For $i,j=1,2,\dots,N$, denote 
\begin{equation}\label{Denote}
\begin{aligned}
&f_i(t): = g(x_i(t)),~t\in[-\epsilon_i+ s_i, \epsilon_i + s_i],\\
&r_{ij} (t,s) := R_{\partial\Omega _h}\left( x_i(t),x_j(s) \right), ~t \in [-\epsilon_i+ s_i,\epsilon_i+ s_i], s \in [-\epsilon_j+ s_j,\epsilon_j+ s_j],\\
&\phi_i(t): = \frac{\partial u(x_i(t))}{\partial \nu},~t\in[-\epsilon_i+ s_i, \epsilon_i + s_i].
\end{aligned}
\end{equation}
It then follows from \eqref{u_original} that 
\beq\label{int_several}
\begin{aligned}
\frac{L_{j}^{2}}{2}-L_j \phi _j \left( t \right) &=f_j(t) -\frac{1}{\pi}\sum_{i=1}^N{\int_{s_i-\epsilon _i}^{s_i+\epsilon _i}{\ln \left| x_j\left( t \right) -x_i\left( s \right) \right|\phi_i \left( s \right) d\sigma \left( s \right)}} \\
&+\sum_{i=1}^N{\int_{s_i-\epsilon _i}^{s_i +\epsilon _i} r_{ji} (t,s)\phi_i \left( s \right) d\sigma \left( s \right)}+C_{\epsilon},
\end{aligned}
\eeq
where $t\in[-\epsilon_j + s_j, \epsilon_j + s_j]$, $j =1,2,\dots,N$.

By changes of variables $t\rightarrow s_i+\epsilon _it$ and $s\rightarrow s_i+\epsilon _is$ for $t,s\in(s_i-\epsilon_i,s_i+\epsilon_i)$, the above integral equation \eqref{int_several} becomes
\beq\label{matrix_several}
\left\{ \begin{array}{l}
	\displaystyle\frac{L_{1}^{2}}{2}-\frac{L_1}{\epsilon _1}\phi _1\left( t \right) =f_1(s_1+\epsilon_1 t)-\frac{1}{\pi}\sum\limits_{i=1}^N{\int_{-1}^1{\ln \left| x_1\left( s_1+\epsilon _1t \right) -x_i\left( s_i+\epsilon _is \right) \right|\phi _i\left( s \right) ds}}\\
	\displaystyle \hspace{4cm}+\sum\limits_{i=1}^N{\int_{-1}^1{ r_{1i} (x_1\left( s_1+\epsilon _1t \right) , x_i\left( s_i+\epsilon _is \right)) \phi _i\left( s \right) ds}}+C_{\epsilon},\\
	\displaystyle\frac{L_{2}^{2}}{2}-\frac{L_2}{\epsilon _2}\phi _2\left( t \right) =f_2(s_2+\epsilon_2 t)-\frac{1}{\pi}\sum\limits_{i=1}^N{\int_{-1}^1{\ln \left| x_2\left( s_2+\epsilon _2t \right) -x_i\left( s_i+\epsilon _is \right) \right|\phi _i\left( s \right) ds}}\\
	\displaystyle \hspace{4cm}+\sum\limits_{i=1}^N{\int_{-1}^1{ r_{2i} (x_2\left( s_2+\epsilon _2t \right) , x_i\left( s_i+\epsilon _is \right)) \phi _i\left( s \right) ds}}+C_{\epsilon},\\
	\hspace{3cm} \vdots\\
	\displaystyle\frac{L_{N}^{2}}{2}-\frac{L_N}{\epsilon _N}\phi _N\left( t \right) =f_N(s_N+\epsilon_N t)-\frac{1}{\pi}\sum\limits_{i=1}^N{\int_{-1}^1{\ln \left| x_N\left( s_N+\epsilon _Nt \right) -x_i\left( s_i+\epsilon _is \right) \right|\phi _i\left( s \right) ds}}\\
	\displaystyle \hspace{4cm}+\sum\limits_{i=1}^N{\int_{-1}^1{ r_{Ni} (x_N\left( s_N+\epsilon _N t \right) , x_i\left( s_i+\epsilon _is \right)) \phi _i\left( s \right) ds}}+C_{\epsilon},\\
\end{array} \right. 
\eeq
where $\phi _i\left( t \right) =\epsilon _i\phi_i \left( s_i+\epsilon _it \right) $ for $t\in[-1,1]$.

Define integral operators $P,P_j: L^2\left[ -1,1 \right] \rightarrow L^2\left[ -1,1 \right] 
$ as
\begin{equation*}
\begin{aligned}
P\left[ \psi \right] &:=\int_{-1}^1{\ln \left| t-s \right|\psi \left( s \right) ds};\\
P_j \left[ \psi \right] &:=\frac{1}{\epsilon_j}\int_{-1}^1\big\{\ln \frac{\left| x_j\left( s_j+\epsilon _j t \right) -x_j\left( s_j+\epsilon _j s \right) \right|}{\epsilon_j \left| t-s \right|}
+\pi r_{jj}\left( s_j,s_j \right)\\
& \hspace{3cm}  -\pi r_{jj}\left( s_j+\epsilon _j t, s_j +\epsilon _j s \right) \big\} \psi \left( s \right) ds,\quad j = 1,\dots,N.
\end{aligned}
\end{equation*}
One can easily check that 
\begin{equation*}
\left\{ \begin{array}{l}
	\ln \left| x_j \left( s_j+\epsilon _j t \right) -x_j \left( s_j +\epsilon _j s \right) \right|=\ln \epsilon _j \left| t-s \right|+O\left( \epsilon _j \right) ,\\
	\ln \left| x_i\left( s_i+\epsilon _it \right) -x_j\left( s_j+\epsilon _js \right) \right|=\ln \left| s_i -s_j \right|+O\left( \sqrt{\epsilon _{i}^{2}+\epsilon _{j}^{2}} \right) ,\\
\end{array} \right. 
\end{equation*}
for $i,j=1,2,...,N;\; i \ne j$.  Thus the operators $P$ and $P_j$ are bounded independently of $\epsilon_j$, $j = 1,2,\dots,N$.

Integrating the first equation in \eqref{NR} over $\Omega _h$ and using the divergence theorem, we obtain the compatibility condition
\beq\label{compatibility_o}
\sum_{i = 1}^N \int_{\Gamma_{\epsilon_i}}{\partial _{\nu}ud\sigma} = -|\Omega _h|.
\eeq
Let
\beq\label{Ci}
	C_i=\int_{-1}^1{\phi_i}\left( t \right) dt,\quad i=1,2,\dots,N.
\eeq
By \eqref{compatibility_o} and the third equation of \eqref{Denote}, we have
\beq\label{compatibility}
	\sum_{i = 1}^N  C_i =-|\Omega _h|.
\eeq
Thus we can rewrite \eqref{matrix_several} as
\beq\label{C_eps}
\left\{ \begin{array}{l}
	\left( I-\dfrac{\epsilon _1}{\pi L_1}\left( P+\epsilon_1 P_1 \right) \right) \phi _1\left( t \right) =\epsilon _1{C}_{\epsilon}^{1}+O\left( \epsilon _1 \tilde{\epsilon} \right),\\[1em]
	\left( I-\dfrac{\epsilon _2}{\pi L_2}\left( P+\epsilon_2 P_2 \right) \right) \phi _2\left( t \right) =\epsilon _2 {C}_{\epsilon}^{2}+O\left( \epsilon _2 \tilde{\epsilon} \right),\\
	\hspace{5cm} \vdots\\
	\left( I-\dfrac{\epsilon _N}{\pi L_N}\left( P+\epsilon_N P_N \right) \right) \phi _N\left( t \right) =\epsilon _N{C}_{\epsilon}^{N}+O\left( \epsilon _N \tilde{\epsilon} \right),
\end{array} \right. 
\eeq
where
\beq\label{Ce}
\begin{aligned}
	{C}_{\epsilon}^{i}=\frac{L_i}{2}-\frac{1}{L_i}\left( f_i\left( s_i \right) + \sum_{j= 1}^N C_j r_{ij} +C_{\epsilon}\right)+\frac{1}{\pi L_i} \sum_{j \neq i} C_j \ln \left|  s_i -s_j \right|+\frac{C_i}{\pi L_i}\ln \epsilon _i,
\end{aligned}
\eeq
and $\tilde{\epsilon} = \max \{\epsilon_1,\epsilon_2,\dots,\epsilon_N \}$, $r_{ij}=r_{ij}(s_i,s_j)$ for $i,j=1,2,\dots,N$. Note that  $r_{ij}=r_{ji}$, when $i\ne j$.

Since we assume that $\frac{\epsilon _i}{\pi L_i}\ll1 $ for $i=1, \dots, N$, one can easily see that
\beq\label{inv}
\left( I-\frac{\epsilon _i}{\pi L_i}\left( P +\epsilon _i P_i \right) \right) ^{-1}=I+\frac{\epsilon _i}{\pi L_i}\left( P +\epsilon _i P_i \right) +O\left( \epsilon _{i}^{2} \right).
\eeq
Thus, by \eqref{C_eps} and \eqref{inv}, we have
\beq\label{phi}
	\phi _i \left( t \right) =\epsilon _i { C}_{\epsilon}^{i}\left(1 +\dfrac{\epsilon _{i}}{\pi L_i}P \left[ 1 \right] \left( t \right) \right)+O\left( \epsilon _i \tilde{\epsilon} \right).
\eeq

Integrating \eqref{phi} over $(-1,1)$ and by \eqref{Ci}, we have 
\beq\label{C_e}
	\dfrac{C_i}{2\epsilon _i}={C}_{\epsilon}^{i}\left( 1+\dfrac{\epsilon _i}{2\pi L_i}\int_{-1}^1{P \left[ 1 \right]}\left( t \right) dt \right) +O\left(\tilde{\epsilon}\right).
\eeq
It is easy to see that
\begin{equation*}
\left( 1+\frac{\epsilon _i}{2\pi L_i}\int_{-1}^1{P \left[ 1 \right]}\left( t \right) dt \right) ^{-1}=1-\frac{\epsilon _i}{2\pi L_i}\int_{-1}^1{P \left[ 1 \right]}\left( t \right) dt+O\left( \epsilon^2 _i \right).
\end{equation*}
Hence, by \eqref{L1} and \eqref{C_e} we have
\beq\label{Cie}
{C}_{\epsilon}^{i}=\frac{C_i}{2\epsilon _i}-\frac{C_i}{2\pi L_i}\left( 2\ln 2-3 \right) +O\left( \tilde{\epsilon} \right) ,\quad i=1, 2,\dots,N.
\eeq

Comparing \eqref{Cie} with \eqref{Ce}, together with \eqref{compatibility}, we obtain the system of $C_1, \dots,C_N$ and $C_\epsilon$:
\beq\label{Cs}
\mathcal{K}\left[ \begin{array}{c}
	C_1\\
	\vdots\\
	C_N\\
	C_\epsilon
\end{array} \right] =\left[ \begin{array}{c}
  \frac{L_{1}^{2}}{2}+f_1\left( s_1 \right) +O\left( \tilde{\epsilon} \right)\\
  \vdots\\
  \frac{L_{N}^{2}}{2}+f_N\left( s_N \right) +O\left( \tilde{\epsilon} \right)\\
  -\left| \Omega _h \right|\\
\end{array} \right],
\eeq
where
$$
\mathcal{K}=\left[ \begin{matrix}
  A_{\epsilon}^{1}&   \frac{-\ln \left| s_1-s_2 \right|}{\pi}+r_{12}&   \frac{-\ln \left| s_1-s_3 \right|}{\pi}+r_{13}&   \cdots&   \frac{-\ln \left| s_1-s_N \right|}{\pi}+r_{1N}&   1\\
  \frac{-\ln \left| s_2-s_1 \right|}{\pi}+r_{12}&   A_{\epsilon}^{2}&   \frac{-\ln \left| s_2-s_3 \right|}{\pi}+r_{23}&   ...&    \frac{-\ln \left| s_2-s_N \right|}{\pi}+r_{2N}&   1\\
  \frac{-\ln \left| s_3-s_1 \right|}{\pi}+r_{13}&   \frac{-\ln \left| s_3-s_2 \right|}{\pi}+r_{23}&   A_{\epsilon}^{3}&   \cdots&   \frac{-\ln \left| s_3-s_N \right|}{\pi}+r_{3N}&   1\\
  \vdots&   \vdots&   \vdots&   \ddots&   \vdots&   \vdots\\
  \frac{-\ln \left| s_N-s_1 \right|}{\pi}+r_{1N}&   \frac{-\ln \left| s_N-s_2 \right|}{\pi}+r_{2N}&   \frac{-\ln \left| s_N-s_3 \right|}{\pi}+r_{3N}&   \cdots&   A_{\epsilon}^{N}&   1\\
  1&    1&    1&    \cdots&   1&    0\\
\end{matrix} \right]
$$
and $A_{\epsilon}^{i}=\frac{L_i}{2\epsilon _i}-\frac{\ln \epsilon _i}{\pi}-\frac{2\ln 2-3}{2\pi}+r_{ii}$.
Solving \eqref{Cs}, we obtain 
$$
C_i = \frac{-\left| \Omega _h \right|\epsilon _i/L_i}{\sum\limits_{k=1}^N{\epsilon _k/L_k}} + O(\tilde{\epsilon} \ln\epsilon_i),
$$
and
\beq\label{Ceps}
C_{\epsilon}=\frac{\left| \Omega _h \right|}{2\sum_{i=1}^N{\epsilon _i/L_i}}+\frac{\left| \Omega _h \right|}{\pi}\left( \sum_{i=1}^{N-1}{\sum_{j=i+1}^N{T_{ij}\ln \epsilon _i\epsilon _j}}-\sum_{i=1}^N{F_i\ln \epsilon _i} \right) +O\left( 1 \right),
\eeq
where the constants $T_{ij}$, $F_i$ are bounded independently of $\epsilon_i$ that are given by
\beq\label{TF}
T_{ij}=\frac{\epsilon _i\epsilon _j}{L_iL_j} \Big/ \left( \sum_{k=1}^N{\frac{\epsilon _k}{L_k}} \right) ^2,\quad F_i=\frac{\epsilon _i}{L_i}\Big/\sum_{k=1}^N{\frac{\epsilon _k}{L_k}}.
\eeq
Thus, by \eqref{phi}, \eqref{Cie} and the third equation of \eqref{Denote}, we have
\beq\label{pu}
\frac{\partial u \left( x_i(t) \right)}{\partial \nu}=\frac{C_i}{2\epsilon _i}-\frac{C_i}{2\pi L_i}\left( 2\ln 2-3 \right) +\frac{C_i}{2\pi L_i} P \left[ 1 \right] \left( \frac{t-s_i}{\epsilon _i} \right) +O\left( \tilde{\epsilon} \right) ,
\eeq
where $ t\in\left( s_i-\epsilon _i,s_i+\epsilon _i \right),\; i=1, \dots, N$ .

Finally, substituting \eqref{pu} and \eqref{Ceps} into \eqref{u_multi}, we arrive at the high order asymptotic expansion of the MFPT $u$: 
\beq\label{uO1}
u(x)=\frac{\left| \Omega _h \right|}{2\sum_{i=1}^N{\epsilon _i/L_i}}+\frac{\left| \Omega _h \right|}{\pi}\left( \sum_{i=1}^{N-1}{\sum_{j=i+1}^N{T_{ij}\ln \epsilon _i\epsilon _j}}-\sum_{i=1}^N{F_i\ln \epsilon _i} \right)+O\left( 1 \right),
\eeq
where $T_{ij}$, $F_i$ are given by \eqref{TF}. This is the end of the proof of Theorem 1. 

If $\Omega_h$ is a general shaped $C^2$-boundary domain, $N$ necks have the same length $L$ and the same width $2\epsilon$, then the MFPT $u$ has the following explicit three-term asymptotic expension up to accuracy $O\left( \epsilon \ln ^2\epsilon \right)$:
\beq\label{u_N}
\begin{aligned}
u\left( x \right) =&\frac{\left| \Omega _h \right|L}{2N}\frac{1}{\epsilon}-\frac{\left| \Omega _h \right|}{\pi N}\ln \epsilon -\frac{\left| \Omega _h \right|\left( 2\ln 2-3 \right)}{2\pi N}+\frac{L^{2}}{2}-\frac{2\left| \Omega _h \right|}{\pi N^2}\sum_{i=1}^{N-1}\sum_{j=i+1}^N\ln \left|s_i-s_j \right|\\
&+\int_{\Omega _h}{G \left( x,y \right) dy}-\frac{1}{N}\sum\limits_{i=1}^{N}\int_{\Omega _h}{G \left( s_i,y \right) dy}-\frac{\left| \Omega _h \right|}{N}\sum_{i=1}^N{G_{\partial \Omega _h}\left( x,s_i \right)}\\
&+\frac{\left| \Omega _h \right|}{N^2}\sum\limits_{i=1}^N{r_{ii}}+\frac{2\left| \Omega _h \right|}{N^2}\sum\limits_{i=1}^{N-1}{\sum_{j=i+1}^N{r_{ij}}}+O\left( \epsilon \ln ^2\epsilon \right) .
\end{aligned}
\eeq
As mentioned in the Introduction, the first leading term of \eqref{u_N} is proportional to the length $L$ and of order $O(1/\epsilon)$, which is different from the well-known leading term $O(\ln \epsilon)$ for two dimensional NEP without necks. Due to the existence of narrow necks, it takes longer time for the particle to escape from the neck gate, which is natural and interesting. 

Furthermore, if we assume that $\Omega_h$ is a unit disk centered at $(0,0)$, then $R_{\partial \Omega_h}\left( x,y \right)=0 $ and $
\int_{\Omega _h}{G\left( x,y \right) dy=0}$ for $x\in \partial \Omega_h$. In this way, \eqref{u_N} is reduced to the following simple form:
\beq\label{u_N}
\begin{aligned}
u\left( x \right) =&\frac{\left| \Omega _h \right|L}{2N}\frac{1}{\epsilon}-\frac{\left| \Omega _h \right|}{\pi N}\ln \epsilon -\frac{\left| \Omega _h \right|}{2\pi N}\left( 2\ln 2-3 \right)+\frac{L^{2}}{2}-\frac{2\left| \Omega _h \right|}{\pi N^2}\sum_{i=1}^{N-1}\sum_{j=i+1}^N \ln \left| s_i-s_j \right|\\
&+\frac{1}{4}\left( 1-\left| x \right|^2 \right) +\frac{\left| \Omega _h \right|}{\pi N}\sum_{i=1}^N{\ln \left| x-s_i  \right|}+O\left( \epsilon \ln ^2\epsilon \right) .
\end{aligned}
\eeq
As it is also mentioned in the Introduction, the first and second leading order terms of the above formula is $1/N$ of \eqref{u1} which is derived in \cite{Li}, where there exists only a single neck. This is  is quite natural. However, the third leading order term $O(1)$ does not satisfy the $1/N$ relation, since $O(1)$ term depends not only on the location of the starting point, the length of the neck, but also on the location of the narrow necks as well as the interaction between them. The importance of the role that the $O(1)$ term plays to the accuracy of the MFPT is shown through the numerical results in Section \ref{Numerical}. 

Moreover, suppose that only two necks $\Omega_{n_1}$ and $\Omega_{n_2}$ are connected to a general shaped domain $\Omega_h$ with $C^2$ boundary, with different length $L_1$ and $L_2$, and different width $2\epsilon_1$ and $2\epsilon_2$, respectively. Then the MFPT \eqref{uO1} has the following three-term asymptotic expansion by considering different $L_i$ and $\epsilon_i$: 
\beq\label{u2}
\begin{aligned}
u \left( x \right) =&\frac{\left| \Omega _h \right|}{2\left( \frac{\epsilon _1}{L_1}+\frac{\epsilon _2}{L_2} \right)}+\frac{\left| \Omega _h \right|}{\pi}\left[ \left( T-F_1 \right) \ln \epsilon _1+\left( T-F_2 \right) \ln \epsilon _2 \right] +\mathcal{Q}\left( x \right)+C \\
& \hspace{3cm} +O\left( \sqrt{\epsilon _{1}^{2}+\epsilon _{2}^{2}}\ln \epsilon _1\ln \epsilon _2\right),
\end{aligned}
\eeq
where $T$, $F_1$, $F_2$ and $C$ are bounded independently of $\epsilon_1$ and $\epsilon_2$, that are given by
$$
T=\dfrac{\epsilon _1\epsilon _2}{L_1L_2 \left( \frac{\epsilon _1}{L_1}+\frac{\epsilon _2}{L_2} \right) ^2}, \quad F_1=\frac{\epsilon _1}{L_1\left( \frac{\epsilon _1}{L_1}+\frac{\epsilon _2}{L_2} \right)}, \quad F_2=\frac{\epsilon _2}{L_2 \left( \frac{\epsilon _1}{L_1}+\frac{\epsilon _2}{L_2} \right) }
$$
and
$$
\begin{aligned}
C&=-\left( \sum_{i=1}^2{\sum_{j=1}^2{\left( -1 \right) ^{i+j}r_{ij}+\frac{1}{\pi}\left( 2\ln \left| s_1 -s_2 \right|-2\ln 2+3 \right)}} \right) T\left| \Omega _h \right|\\
&+\sum_{i=1}^2{\left( \left| \Omega _h \right|F_i\left( -\frac{2\ln 2-3}{2\pi}+r_{ii} \right) \right) +\sum_{i=1}^2{F_i}\left( \frac{L_{i}^{2}}{2}-f_i\left( s_i \right) \right)},
\end{aligned}
$$
where $s_1$ and $s_2$ are the center point of $\Gamma_{\epsilon_1}$ and $\Gamma_{\epsilon_2}$, respectively. Here the $x$-dependent term $Q(x)$ is given by
$$
\mathcal{Q}\left( x \right)=\int_{\Omega _h}{G\left( x,y \right) dy}-\frac{\epsilon _1L_2\left| \Omega _h \right|}{\epsilon _2L_1+\epsilon _1L_2} G_{\partial \Omega _h}\left( x,s_1 \right) -\frac{\epsilon _2L_1\left| \Omega _h \right|}{\epsilon _2L_1+\epsilon _1L_2} G_{\partial \Omega _h}\left( x,s_2 \right)
$$
for $x\in\Omega_h$ provided that $dist(x,\Gamma_{\epsilon_i})>c_0$, for some constant $c_0> 0$. One can easily see that $\mathcal{Q}\left( x \right)$ is bounded for $x\in\Omega_h$. The asymptotic expansion \eqref{u2} gives explicit formula of the three-terms $O(1/\epsilon)$, $O(\ln \epsilon)$ and $O(1)$, for different lengths and different widths of the necks. If $\Omega_h$ is a unit disk, and $\epsilon_1=\epsilon_2=\epsilon$, $L_1=L_2=L$, then the MFPT \eqref{u2} is reduced to the following simple form:
\beq\label{two}
\begin{aligned}
u\left( x \right) =&\frac{\pi L}{4}\frac{1}{\epsilon}-\frac{1}{2}\ln \epsilon -\frac{1}{4}\left( 2\ln 2-3 \right)+\frac{L^{2}}{2}-\frac{1}{2}\ln\left| s_1 -s_2\right|\\
&+\frac{1}{4}(1-|x|^2) + \frac{1}{2}\left(\ln\left| x-s_1 \right| + \ln\left| x-s_2 \right| \right)+O(\epsilon\ln^2\epsilon).
\end{aligned}
\eeq
It is worth mentioning that \eqref{two} is only valid when two necks are well-separated, i.e., $|s_2-s_1|\geq c$, for some $c>0$. 

In the rest of this section, we consider the case when two necks are not well-separated, i.e., $|s_2-s_1| = d\epsilon$, for $d>2$. We still assume that $\Omega_h$ is a unit disk and $\epsilon_1=\epsilon_2=\epsilon$, $L_1=L_2=L$. Then the system \eqref{matrix_several} becomes
\beq\label{c2}
\left\{ \begin{array}{l}
	\displaystyle\frac{L^{2}}{2}-\frac{L}{\epsilon}\phi _1\left( t \right) =-\frac{1}{\pi}{\int_{-1}^1{\ln \left| x_1\left( s_1+\epsilon t \right) -x_1\left( s_1+\epsilon s \right) \right|\phi _1\left( s \right) ds}}\\
	\displaystyle \hspace{4cm} -\frac{1}{\pi}{\int_{-1}^1{\ln \left| x_1\left( s_1+\epsilon t \right) -x_2\left( s_2+\epsilon s \right) \right|\phi _2\left( s \right) ds}}+C_{\epsilon},\\
	\displaystyle\frac{L^{2}}{2}-\frac{L}{\epsilon }\phi _2\left( t \right) =-\frac{1}{\pi}{\int_{-1}^1{\ln \left| x_2\left( s_2+\epsilon t \right) -x_1\left( s_1+\epsilon s \right) \right|\phi _1\left( s \right) ds}}\\
	\displaystyle \hspace{4cm}-\frac{1}{\pi}{\int_{-1}^1{\ln \left| x_2\left( s_2+\epsilon t \right) -x_2\left( s_2+\epsilon s \right) \right|\phi _2\left( s \right) ds}} + C_{\epsilon},
\end{array} \right. 
\eeq
where $\phi _i\left( t \right) =\epsilon _i\phi_i \left( s_i+\epsilon t \right) $ for $t\in[-1,1]$.
Define
\begin{equation*}
\begin{aligned}
Q_1\left[ \psi \right] (t)&:=\int_{-1}^1{\ln \left| d+t-s \right|\psi \left( s \right) ds},\\
Q_2\left[ \psi \right] (t)&:=\int_{-1}^1{\ln \left| d-t+s \right|\psi \left( s \right) ds},
\end{aligned}
\end{equation*}
then $Q_1$ and $Q_2$ are bounded from $L^2[-1,1]$ to $L^2[-1,1]$. The  system of equations \eqref{c2} can be written as
\beq\label{phi_2}
\left[\begin{matrix}
\displaystyle I-\dfrac{\epsilon}{\pi L} P & -\dfrac{\epsilon}{\pi L}Q_2 \\[1em]
\displaystyle -\dfrac{\epsilon}{\pi L}Q_1 & I-\dfrac{\epsilon}{\pi L}P 
\end{matrix}\right] \left[\begin{matrix} \phi_1 \\[1em] \phi_2 \end{matrix} \right] = \epsilon \left[\begin{matrix}  - \frac{C_{\epsilon} }{L} - \frac{1}{L}\ln \epsilon +\frac{L}{2} \\[1em]  - \frac{C_{\epsilon} }{L} - \frac{1}{L}\ln \epsilon  + \frac{L}{2}\end{matrix} \right] + O(\epsilon^2).
\eeq
It is easy to see that
$$
\left[\begin{matrix}
\displaystyle I-\dfrac{\epsilon}{\pi L} P & -\dfrac{\epsilon}{\pi L}Q_2 \\[1em]
\displaystyle -\dfrac{\epsilon}{\pi L}Q_1 & I-\dfrac{\epsilon}{\pi L}P
\end{matrix}\right] ^{-1} = \left[\begin{matrix}
\displaystyle I + \dfrac{\epsilon}{\pi L} P & \dfrac{\epsilon}{\pi L}Q_2 \\[1em]
\displaystyle \dfrac{\epsilon}{\pi L}Q_1 & I + \dfrac{\epsilon}{\pi L}P
\end{matrix}\right] + O(\epsilon^2).
$$
{\color{blue}Denote}
$$A = \left[\begin{matrix}
P & Q_2 \\
Q_1 & P
\end{matrix}\right]. $$
Then we can solve \eqref{phi_2} as follows
\beq\label{phi12}
 \left[\begin{matrix} \phi_1 \\ \phi_2 \end{matrix} \right]
 = \frac{\epsilon}{L} \left(- C_{\epsilon} - \ln \epsilon + \frac{L^2}{2} -\dfrac{\epsilon}{\pi L} C_{\epsilon} A  \left[\begin{matrix} 1\\1 \end{matrix} \right]\right) + O(\epsilon^2\ln\epsilon).
\eeq
Denote 
$$\left[\begin{matrix} \alpha_1 \\ \alpha_2 \end{matrix} \right]
:= \int_{-1}^{1}A\left[\begin{matrix} 1\\1 \end{matrix} \right](t)dt.$$
Integrating \eqref{phi12} over $(-1,1)$ and by the compatibility condition, we obtain 
\beq\label{C_eps_2}
C_\epsilon = \frac{\pi L}{4\epsilon} - \ln \epsilon + \frac{L^2}{2} -\frac{1}{16}(\alpha_1+\alpha_2) + O(\epsilon\ln\epsilon).
\eeq
Therefore, by \eqref{u_multi}, \eqref{phi12} and \eqref{C_eps_2}, we obtain the following result.
\begin{thm}
Let $\Omega_h$ be the unit disk. Suppose that two necks with the same width $2\epsilon$ and the same length $L$, are not well-separated, i.e., $|s_2-s_1| = d\epsilon$, for $d>2$.  Then the MFPT $u$ to \eqref{original_u} is given asymptotically by
\beq\label{separated}
\begin{aligned}
u(x)&=\frac{\pi L}{4\epsilon} - \ln \epsilon -\frac{1}{16}(\alpha_1+\alpha_2) + \frac{L^2}{2} \\
&+\frac{1}{4}(1-|x|^2) + \frac{1}{2}\left(\ln\left| x-s_1 \right| + \ln\left| x-s_2 \right| \right)+O(\epsilon\ln^2\epsilon).
\end{aligned}
\eeq
\end{thm}
By comparing the formula \eqref{separated} for two clustered necks with \eqref{two} for two well-separated necks, it is interesting to see that the interaction between two necks is described by the term $- \ln \epsilon -\frac{1}{16}(\alpha_1+\alpha_2)$ for the clustered case, while it is described by the term $-\frac{1}{2}\ln|s_1-s_2|$ for well-separated case. As the number $d\rightarrow 2+$, the two clustered necks will be combined into one neck of width $4\epsilon$, in this case, we can calculate that 
\beq\label{alpha}
\begin{aligned}
\alpha_1+\alpha_2 &= \int_{-1}^1{P\left[ 1 \right]}\left( t \right) dt + \int_{-1}^1{Q_2\left[ 1 \right]}\left( t \right) dt + \int_{-1}^1{Q_1\left[ 1 \right]}\left( t \right) dt +\int_{-1}^1{P\left[ 1 \right]}\left( t \right) dt \\
& = 32\ln 2-24.
\end{aligned}
\eeq
Substituting \eqref{alpha} into \eqref{separated} one can see that the solution \eqref{separated} converges to the one \eqref{u1} with a single neck gate of radius $4\epsilon$, i.e.,
$$\frac{\pi L}{4\epsilon} - \ln \epsilon -(2\ln 2-\frac{3}{2})+ \frac{L^2}{2} +\frac{1}{4}(1-|x|^2) + \ln\left| x-s_0 \right|,$$
which is surprising and reasonable.

\section{Numerical experiments}\label{Numerical}

In this section, we conduct numerical experiments to verify the three-term asymptotic expansion  by comparing it with the numerical solution $u_R$ to the original narrow escape problem \eqref{original_u}. Let $u_r$ be the numerical solution to the Neumann-Robin Model \eqref{NR_eq}. We also compare $u_r$ with $u_R$ to show the equivalence of two models. The numerical computations of $u_R$ and $u_r$ are conducted using COMSOL 5.6. The finite element PDE solver is used. In order to obtain more accurate numerical results of $u_R$ and $u_r$, the triangular meshes are locally refined near the connecting parts between the head and necks with element size scaling factor $0.1$.

\begin{figure}[!ht]
\centering
\begin{subfigure}{1\textwidth}
\centering
\includegraphics[width=150mm]{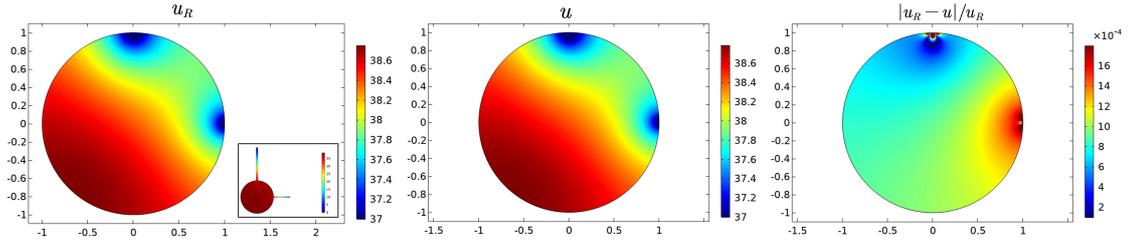}
\caption{$L_1=1,\; L_2=2$, $\epsilon_1=0.02, \;\epsilon_2=0.05$.}
\label{two1}
\end{subfigure}
\begin{subfigure}{1\textwidth}
\centering
\includegraphics[width=150mm]{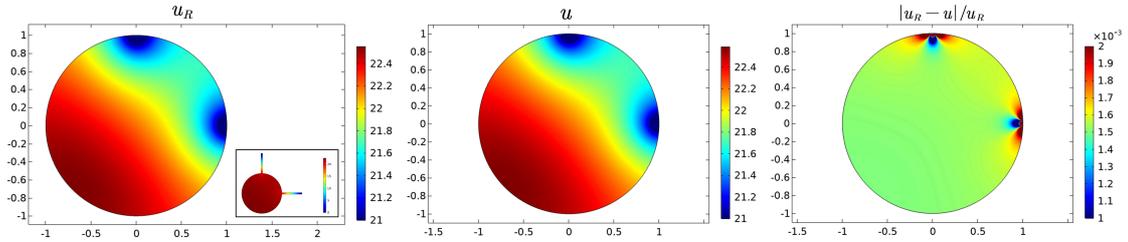}
\caption{$L_1=L_2=1$, $\epsilon_1=\epsilon_2=0.04$.}
\label{two2}
\end{subfigure}
\begin{subfigure}{1\textwidth}
\centering
\includegraphics[width=150mm]{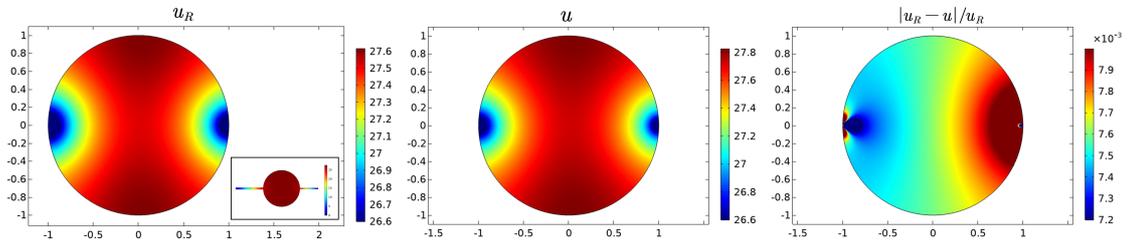}
\caption{$L_1=1,\; L_2=1.5$, $\epsilon_1=0.03, \;\epsilon_2=0.05$.}
\label{two3}
\end{subfigure}
\caption{The numerical solution $u_R$ to original problem (first column), the asymptotic solution $u$ (second column) and the relative error between $u_R$ and $u$ (third column).}
\label{fig:disk1}
\end{figure}

Let $\Omega_h$ be the unit disk centered at $(0,0)$. Figure \ref{fig:disk1} shows the numerical results with two necks. The first column shows the numerical result of $u_R$, the second column shows that of our asymptotic solution $u$, and the third column shows the relative error between them. We consider the following three cases: (i) Let $L_1=1,\; L_2=2$, $\epsilon_1=0.02, \;\epsilon_2=0.05$. Two necks are perpendicularly connected to $\Omega_h$. From the first figure of Figure \ref{two1}, one can see that the MFPT $u_R$ ranges from $37$ to $38.6$ which depends on the location of the starting points. From the asymptotic formula \eqref{u2}, the first two leading order terms of $u$ is $\frac{1}{2}\left| \Omega _h \right| /\left( \frac{\epsilon _1}{L_1}+\frac{\epsilon _2}{L_2} \right)+\frac{\left| \Omega _h \right|}{\pi}\left( \left( T-F_1 \right) \ln \epsilon _1+\left( T-F_2 \right) \ln \epsilon _2 \right)  = 36.6039$, which is close to $u_R$, but not accurate. Thus, the third leading order term $O(1)$ contributes to the accuracy of our asymptotic solution, such that the relative error between $u_R$ and $u$ is small of order $O(10^{-4})$ which can be seen from the third figure of  Figure \ref{two1}. (ii) Let $L_1=L_2=1$, $\epsilon_1=\epsilon_2=0.04$. Two necks are perpendicularly connected to $\Omega_h$. One can see from Figure \ref{two2} that the relative error between $u_R$ and $u$ is small. (iii) Let $L_1=1,\; L_2=1.5$, $\epsilon_1=0.03, \;\epsilon_2=0.05$. Two necks are connected to $\Omega_h$ parallel in the opposite direction. Figure \ref{two3} shows that the relative error is as small as $O(10^{-3})$. In this experiment, by comparing the numerical results with our asymptotic formula \eqref{u2}, one can see that our asymptotic solution can approximate the MFPT accurately. 

\begin{figure}[!ht]
\centering
\begin{subfigure}{1\textwidth}
\centering
\includegraphics[width=150mm]{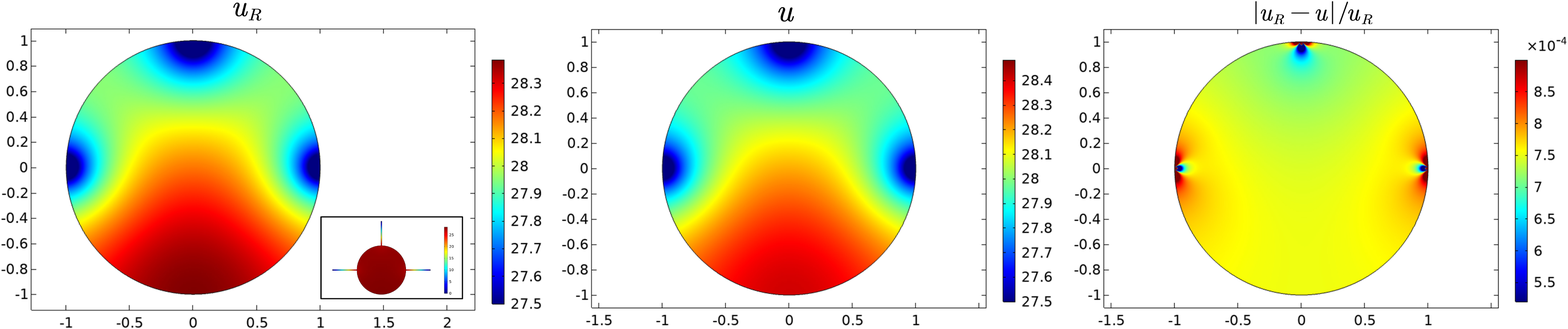}
\caption{Three necks. Neck length: $L_c=1$; Neck width: $2\epsilon=0.04$.}
\label{fig:disk3n}
\end{subfigure}
\begin{subfigure}{1\textwidth}
\centering
\includegraphics[width=150mm]{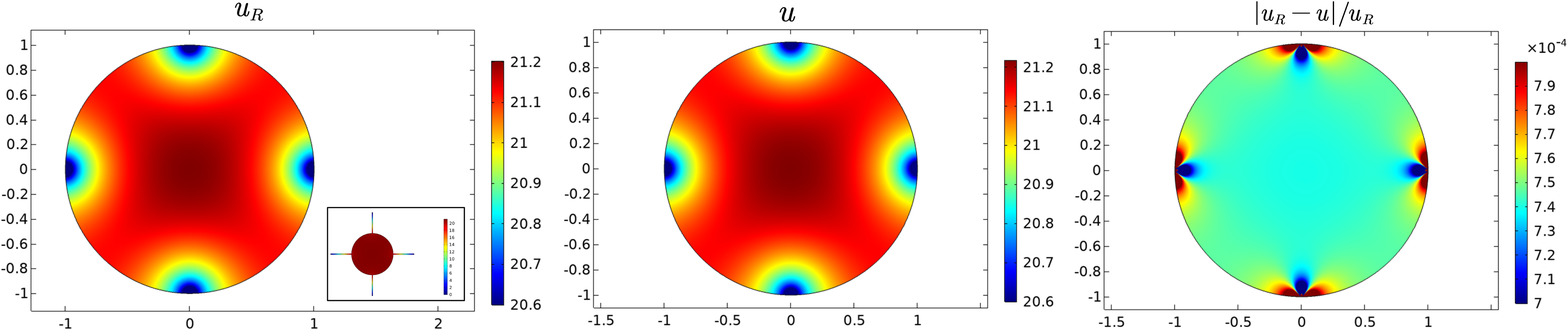}
\caption{Four necks. Neck length: $L_c=1$; Neck width: $2\epsilon=0.04$.}
\label{fig:disk4n}
\end{subfigure}
\caption{The numerical solution $u_R$ to original problem (first column), the asymptotic solution $u$ (second column) and the relative error between $u_R$ and $u$ (third column).}
\label{fig:diskn}
\end{figure}

Let the unit disk be connected with several necks of the same length $L = 1$ and same width $2\epsilon = 0.04$. Figure \ref{fig:disk3n} shows the comparison between $u_R$ and $u$ with three necks. Figure \ref{fig:disk4n} shows that with four necks. The inset of Figure \ref{fig:diskn} shows the location of the necks. One can see from the third column that the relative errors for both three necks and four necks are of order $O(10^{-4})$. 

Furthermore, we fix $\epsilon_1=\epsilon_2=0.01$ and vary the neck lengths $L_1$ and $L_2$ from $1$ to $4$.  For each pair of $L_1$ and $L_2$, we compare the values of the numerical solution $u_R$, $u_r$ and our asymptotic solution $u$ given by \eqref{u2}. We set the particle initiated at the center point $(0,0)$ of the unit disk. Table~\ref{tab_L} lists the value of $u_R$, $u_r$, $u$ as well as the value of the relative error $|u_R-u|/u_R$. One can see that the values are in good agreement for different pairs of neck length $L_1$ and $L_2$. Moreover, in order to see more clearly how the third leading order term $O(1)$ contributes to the accuracy of our asymptotic solution, we list the value of the first two leading order terms and the value of the $O(1)$ term in Table \ref{tab_L}, respectively. From Table \ref{tab_L} one can see that the first two terms $O(1/\epsilon)+O(\ln \epsilon)$ is close to the MFPT $u_R$, but not accurate. After adding the third term $O(1)$, the relative error $|u_R-u|/u_R$ becomes as small as $O(10^{-4})$.
\begin{table}[!ht]
    \centering
    \begin{tabular}{cccccccc}
    \hline
    $L_1$ & $L_2$ & $u_R$ & $u_r$ & $u$ &$O(1/\epsilon)+O(\ln \epsilon)$ & $O(1)$ & $|u_R-u|/u_R$ \\ \hline
        1 & 1 & 81.74653 & 81.81745 & 81.82254 & 80.84240& 0.98014& 0.00093 \\ 
        1 & 1.5 & 97.82933 & 97.89082 & 97.89568 & 96.64247& 1.25321& 0.00068 \\ 
        1 & 2 & 108.76071 & 108.81837 & 108.82240 & 107.27818& 1.54422 & 0.00057 \\ 
        1 & 2.5 & 116.70304 & 116.76017 & 116.76131 & 114.92525& 1.83606 & 0.00050 \\ 
        2 & 1.5 & 138.88679 & 138.97472 & 138.98117 & 136.98926& 1.99191 & 0.00068 \\ 
        2.5 & 2 & 179.75845 & 179.84546 & 179.85120 & 176.86394& 2.98726 & 0.00051 \\ 
        3 & 2.5 & 220.66452 & 220.75059 & 220.75602 & 216.52111& 4.23491 & 0.00041 \\ 
        4 & 3 & 278.02288 & 278.11915 & 278.12086 & 271.62895& 6.49191 & 0.00035 \\ \hline
    \end{tabular}
  \caption{$u_R$: numerical solution to the original problem \eqref{original_u}; $u_r$: numerical solution of the Neumann–Robin model \eqref{NR_eq}; $u$: three-term asymptotic solution; $O(1/\epsilon)+O(\ln \epsilon)$: the first two leading order terms of $u$; $O(1)$: The third term of $u$.}
  \label{tab_L}
\end{table}

Next, we fix the neck length $L_1= 1$ and $L_2=2$ and vary the neck width $\epsilon_1$ and $\epsilon_2$. In the same way with Table \ref{tab_L}, for each pair of $\epsilon_1, \epsilon_2$, we compare the values of $u_R$, $u_r$ and $u$ for a particle initiated at the center of the unit disk. Table~\ref{tab_eps} shows the numerical results of these values. One can see that the solutions are in good agreement. As same as Table \ref{tab_L}, the $O(1)$ term of the asymptotic solution $u$ plays an important role, such that the relative error is as small as  $O(10^{-4})$ for each pair of $\epsilon_1, \epsilon_2$, which can be seen from Table \ref{tab_eps}.
\begin{table}[!ht]
      \centering
    \begin{tabular}{cccccccc}
    \hline
         $\epsilon_1$ & $\epsilon_2$ & $u_R$ & $u_r$ & $u$ & $O(1/\epsilon)+O(\ln \epsilon)$ & $O(1)$ & $|u_R-u|/u_R$\\  \hline
        0.028 & 0.028 & 40.88346 & 40.92875 & 40.93055 & 39.38633& 1.54422 &0.00115 \\ 
        0.025 & 0.025 & 45.43363 & 45.47934 & 45.48150 & 43.93728& 1.54422 &0.00105 \\ 
        0.022 & 0.022 & 51.21565 & 51.26125 & 51.26450 & 49.72028& 1.54422 &0.00095 \\ 
        0.019 & 0.019 & 58.81201 & 58.85802 & 58.86172 & 57.31750& 1.54422 &0.00085 \\ 
        0.016 & 0.016 & 69.23454 & 69.28743 & 69.29138 & 67.74716& 1.54422 &0.00082 \\ 
        0.013 & 0.013 & 84.45024 & 84.50652 & 84.51055 & 82.96633& 1.54422 &0.00071 \\ 
        0.010 & 0.010 & 108.76071 & 108.81837 & 108.82240 & 107.27818& 1.54422 &0.00057 \\ 
        0.010  & 0.050 & 48.86885 & 48.92416 & 48.94176 & 46.78426&  2.15749 & 0.00149 \\ 
        0.010  & 0.030 & 66.67597 & 66.72835 & 66.73425 & 64.83104&  1.90321 &  0.00087 \\ 
        0.010  & 0.020 & 82.34406 & 82.39604 & 82.39925 & 80.66911&  1.73014 & 0.00067 \\ \hline
    \end{tabular}
     \caption{$u_R$: numerical solution to the original problem \eqref{original_u}; $u_r$: numerical solution of the Neumann–Robin model \eqref{NR_eq}; $u$: three-term asymptotic solution; $O(1/\epsilon)+O(\ln \epsilon)$: the first two leading order terms of $u$; $O(1)$: The third term of $u$.}
        \label{tab_eps}
\end{table}

Moreover, as mentioned in the Introduction, the $O(1)$ term depends not only on the length of the neck, but also on the location of the starting point $x$ as well as the interaction between the necks. In order to investigate this, let $\Omega_h$ be the unit disk, fix $L_1=1$, $L_2=1.5$, $\epsilon_1=0.01$ and $\epsilon_2=0.02$. We change the starting position $x$ of the Brownian particle, and also change the distance $|s_1-s_2|$ between two necks. Table \ref{tab_3} shows the numerical solution $u_R$, the value of the first two terms $O(1/\epsilon)+O(\ln \epsilon)$, and the value of three terms $O(1/\epsilon)+O(\ln \epsilon)+O(1)$, respectively, for different $x$ and different $|s_1-s_2|$. From Table \ref{tab_3} one can see that the value of the first two-term $O(1/\epsilon)+O(\ln \epsilon)$ stays constant independently on $x$ and $|s_1-s_2|$. The value is close to $u_R$, but not with high accuracy. While the value of the three-term expansion varies along with the starting point $x$ as well as the distance $|s_1-s_2|$.  By taking the $O(1)$ term into account, the relative error $|u_R-u|/u_R$ is as small as $O(10^{-4})$.
\begin{table}[!ht]
    \centering
{
    \begin{tabular}{cccccc}
    \hline
    $x$ & $\left|s_1-s_2\right|$ & two terms & three terms & $u_R$ & $|u_R-u|/u_R$ \\ \hline
        (0,0) & 2 & 69.44308 & 70.62238 & 70.58857 & 0.00048\\ 
        (-0.3,0.4) & 2 & 69.44308 & 70.56863 & 70.53678 & 0.00045 \\
        (0.5,0.35) & 2 & 69.44308 & 70.56449 & 70.52715 & 0.00053 \\
        (-0.25,-0.5) & 2 & 69.44308 & 70.61237 & 70.58006 & 0.00046 \\ 
        (0.63,-0.1) & 2 & 69.44308 & 70.38992 & 70.35063 & 0.00056 \\ 
        (0,0) & $\sqrt{2}$ & 69.44308 & 70.79214 & 70.75748 & 0.00049 \\ 
        (-0.3,0.4) & $\sqrt{2}$ & 69.44308 & 70.63332 & 70.60121 & 0.00045 \\ 
        (0.5,0.35) & $\sqrt{2}$ & 69.44308 & 70.37404 & 70.33829 & 0.00051 \\ 
        (-0.25,-0.5) & $\sqrt{2}$ & 69.44308 & 71.08097 & 71.04586 & 0.00049 \\ 
        (0.63,-0.1) & $\sqrt{2}$ & 69.44308 & 70.41493 & 70.37594 & 0.00055 \\ 
       \hline
    \end{tabular}
  }
  \caption{$x$: starting position of the Brownian particle; $|s_1-s_2|$: distance between two necks; two terms: $O(1/\epsilon)+O(\ln \epsilon)$ of $u$; three terms ($u$): $O(1/\epsilon)+O(\ln \epsilon)+O(1)$; $u_R$: the numerical solution to the original problem \eqref{original_u}.}
  \label{tab_3}
\end{table}

To investigate the role played by the distance between the narrow necks, we perform the following experiment for a Brownian particle initiated at the center of the unit disk. Let $L_1 = 1$, $L_2 = 2$, $\epsilon_1 = 0.01$ and $\epsilon_2 = 0.02$. We fix the position of one neck at position $(1,0)$ and vary the position of the other neck. Table \ref{tab_distance} lists the time for different distance between two necks. From Table \ref{tab_distance} one can see that the MFPT increases as the distance between two necks becomes small.

\begin{table}[!h]
    \centering
   {
    \begin{tabular}{cccccc}
    \hline
     $s_1$  & $s_2$ &$\left|s_1-s_2\right|$ &  $u_R$ & $u$ & $|u_R-u|/u_R$ \\ \hline
     $\left(1,0\right)$ & $\left(-1,0\right)$& 2   & 82.15348   & 82.22597 & 0.00088 \\ 
     $\left(1,0\right)$ & $\left(0,1\right)$ &1.4142 &  82.34406 & 82.39925 & 0.00067   \\
     $\left(1,0\right)$ & $\left(0.837,0.547\right)$ &0.5708 &  82.89891 & 82.85287 & 0.00055  \\
     $\left(1,0\right)$ & $\left(0.994,0.111\right)$ &0.1109 &  83.61491 & 83.67101 & 0.00067   \\ 
     $\left(1,0\right)$ & $\left(0.997,0.071\right)$ & 0.0710 &  83.81951 & 83.89439 & 0.00089  \\ 
     $\left(1,0\right)$ & $\left(0.999,0.033\right)$ & 0.0330 &  84.29013 & 84.27771  & 0.00015  \\ 
       \hline
    \end{tabular}
  }
  \caption{$s_1$: position of one neck; $s_2$: position of the other neck; $u_R$: numerical solution to the original problem \eqref{original_u};  $u$: asymptotic solution. }
   \label{tab_distance}
\end{table}

\begin{figure}[!ht]
  \centering
  \begin{subfigure}{1\textwidth}
  \centering
  \includegraphics[width=70mm]{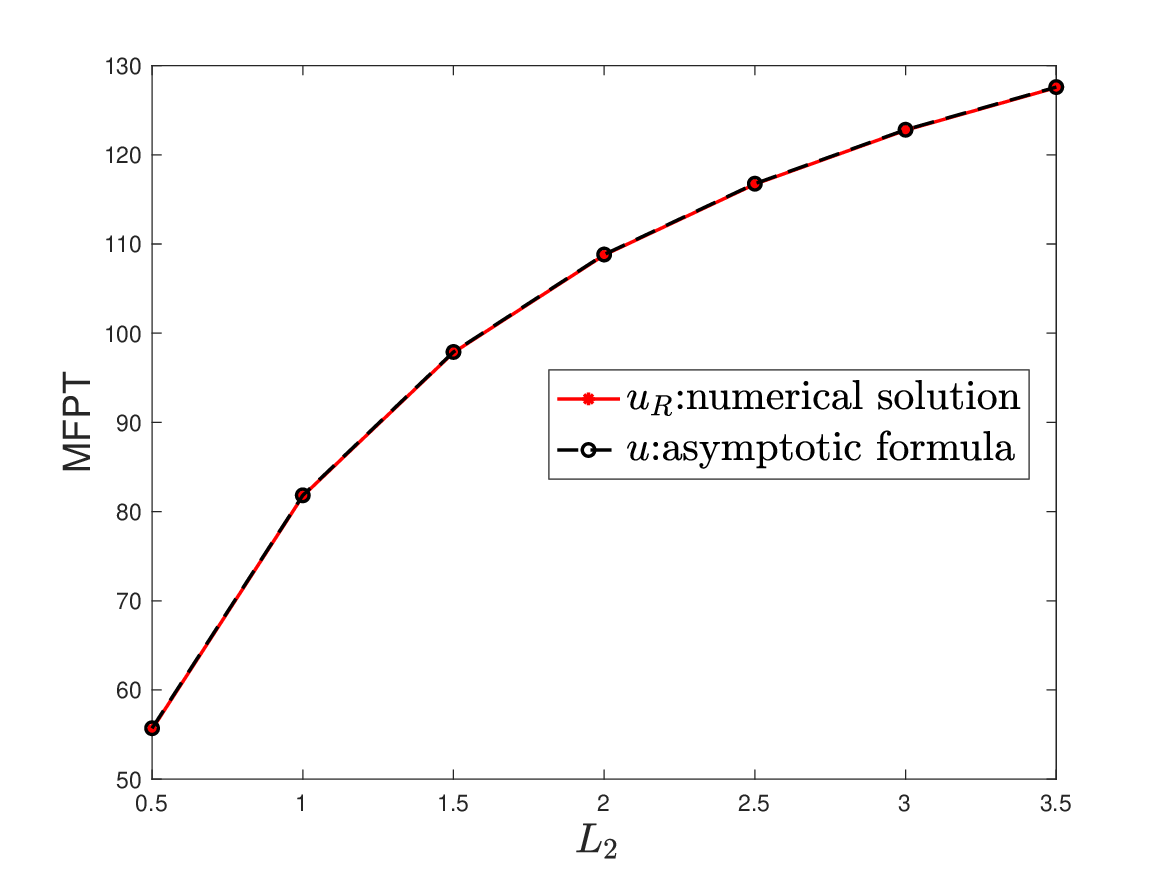}
  \includegraphics[width=70mm]{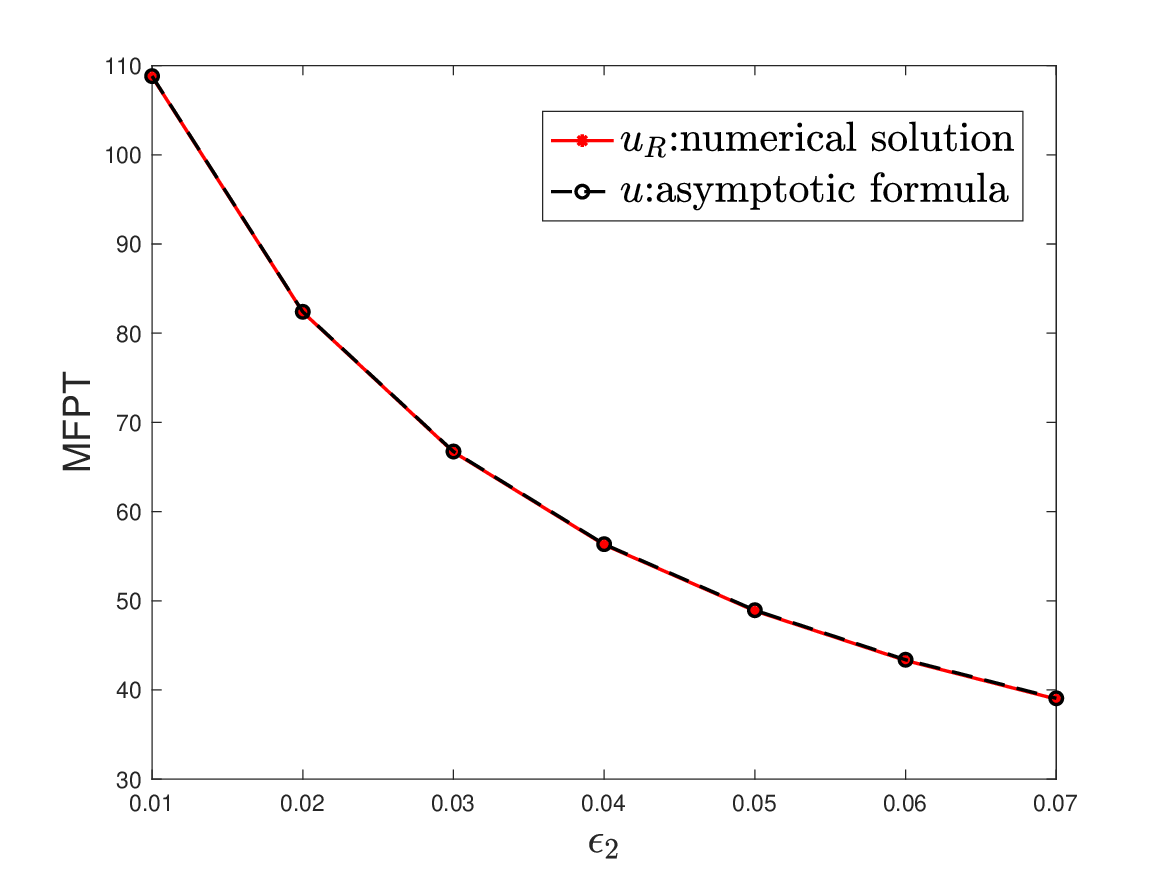}
  \end{subfigure}
  \caption{$u_R$ and $u$ for a particle initiated at the center of the unit disk, with different $L_2$ (left figure) and different neck length $\epsilon_2$ (right figure).}
  \label{fig:curve}
\end{figure}

To see more clearly how our asymptotic formula reveals the escape dynamics and its underlying mechanisms, we perform the following two experiments for a Brownian particle initiated at the center of the unit disk.  (i).  Firstly, we fix $L_1=1$, $\epsilon_1=\epsilon_2=0.01$ and vary the value of $L_2$ from $0.5$ to $3.5$. Figure \ref{fig:curve} shows that our asymptotic formula $u$ matches the MFPT $u_R$ very well. More importantly, one can see from Figure \ref{fig:curve} that as the length $L_2$ increases, the escape time increases as well. The first order term of the formula \eqref{u2} shows the linear dependence on $L_2$. However, the neck length $L_2$ plays role not only in the first term, but also in $O(\ln\epsilon)$ and $O(1)$ term. By taking derivative with respect to $L_2$ for formula \eqref{u2}, the derivation is negative for relatively small value of $\epsilon/L_2 \ll 1$, which is illustrated by the left curve of Figure \ref{fig:curve}. (ii). Secondly, we fix $L_1=1$, $L_2=2$, $\epsilon_1=0.01$ and vary the value of $\epsilon_2$ from $0.01$ to $0.07$. For small value of $\epsilon_2$, the first term of the formula \eqref{u2} dominates the trend of the escape time. From the curve on the right-hand side of Figure \ref{fig:curve} one can see that the escape time is almost inversely proportional to $\epsilon_2$.

\begin{figure}[!ht]
  \centering
  \includegraphics[width=100mm]{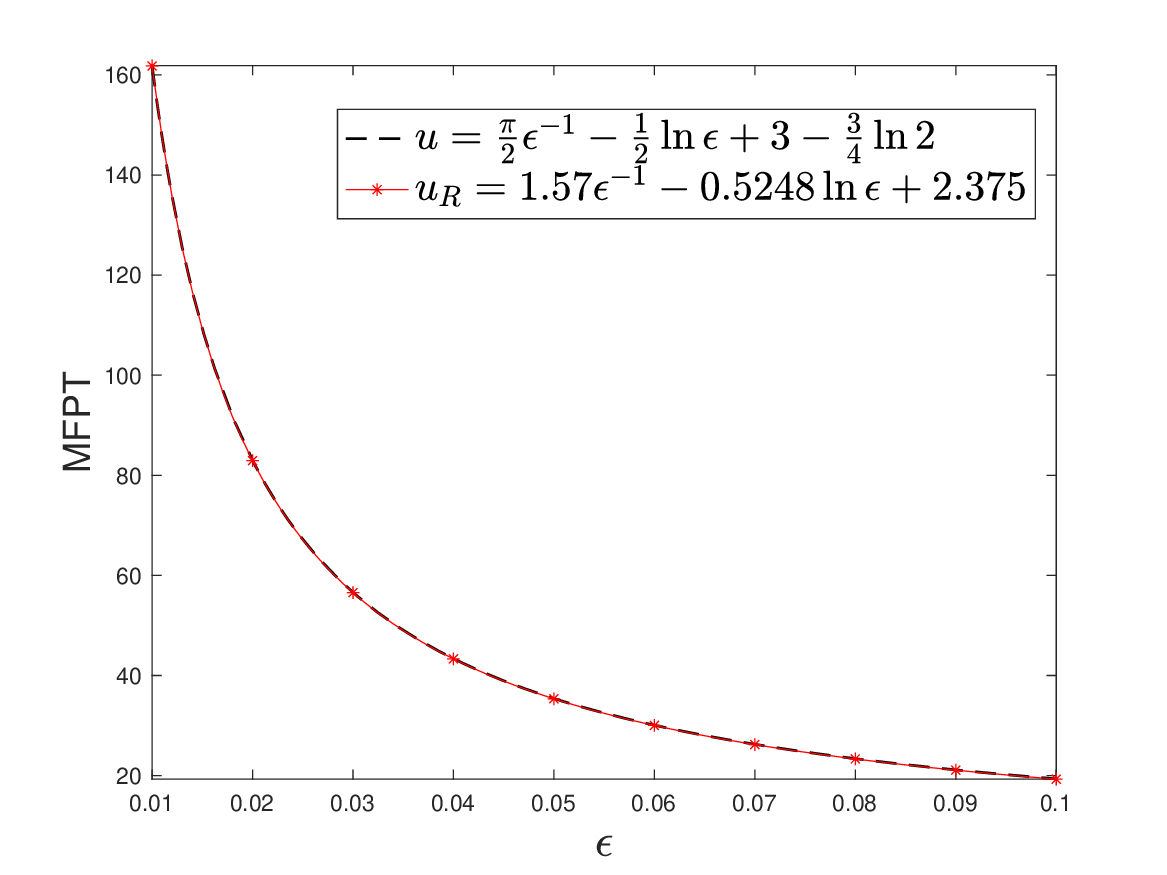}
  \caption{Fitting curve and MFPT for different $\epsilon$.}
  \label{fig:fit}
\end{figure}

Moreover, if the two necks, having the same length $L$ and the same width $2\epsilon$, are perpendicularly connected to $\Omega_h$. For a Brownian particle initiated at the center point $(0,0)$ of $\Omega_h$, the MFPT $u$ has the following form, by \eqref{two}:
\beq\label{u_coef}
u(0,0)=\frac{\pi L}{4}\frac{1}{\epsilon}-\frac{1}{2}\ln \epsilon-\frac{2\ln 2-3}{4}-\frac{\ln \left| s_1 -s_2 \right|}{2}+\frac{L^2}{2}+\frac{1}{4}+O(\epsilon\ln^2\epsilon),
\eeq
where $s_1$ and $s_2$ are the center of the two connecting segments $\Gamma_{\epsilon_1}$ and $\Gamma_{\epsilon_2}$, respectively. In order to confirm the coefficients of the three leading order terms, we perform the following experiment. Fix $L = 2$. From \eqref{u_coef} one can see that the coefficients of the three leading order terms $O(\frac{1}{\epsilon})$, $\ln \epsilon$, $O(1)$ of \eqref{u_coef} are $\pi/2$, $-1/2$ and $3-\frac{3}{4}\ln 2$, respectively.  We now confirm these coefficients by fitting the value of $u_R$ in Table~\ref{tab:fit} with $\epsilon$ decreased from $0.1$ to $0.01$ in a step size of $0.01$. The result is plotted in Figure~\ref{fig:fit}. One can clearly see that the coefficients of the fitting curve $1.57$, $-0.5248$, $2.375$ match well with those of the asymptotic solution $\pi/2$, $-1/2$ and $3-\frac{3}{4}\ln 2$.
\begin{table}[!h]
    \centering
    \begin{tabular}{ccccc}
    \hline
    $\epsilon$ & $u_R$ & $u_r$ & $u$ & $|u_R-u|/u_R$ \\ \hline
        0.10 & 19.28274 & 19.33952 & 19.33940 & 0.00294 \\ 
        0.09 & 21.08308 & 21.13738 & 21.13740 & 0.00258 \\ 
        0.08 & 23.32585 & 23.37776 & 23.37796 & 0.00223 \\ 
        0.07 & 26.20003 & 26.24939 & 26.24972 & 0.00190 \\ 
        0.06 & 30.01901 & 30.06624 & 30.06678 & 0.00159 \\ 
        0.05 & 35.34718 & 35.38037 & 35.39393 & 0.00132 \\ 
        0.04 & 43.31320 & 43.35828 & 43.35949 & 0.00107 \\ 
        0.03 & 56.54429 & 56.59165 & 56.59330 & 0.00087 \\
        0.02 & 82.91620 & 82.97147 & 82.97597 & 0.00072 \\
        0.01 &161.78095 & 161.85738  & 161.8623 & 0.00050\\ \hline
    \end{tabular}
  \caption{$u_R$: numerical solution to the original problem \eqref{original_u}; $u_r$: numerical solution of the Neumann–Robin model \eqref{NR_eq}; $u$: three-term asymptotic solution.}
  \label{tab:fit}
\end{table}

\begin{figure}[!ht]
\centering
\begin{subfigure}{1\textwidth}
\centering
\includegraphics[width=150mm]{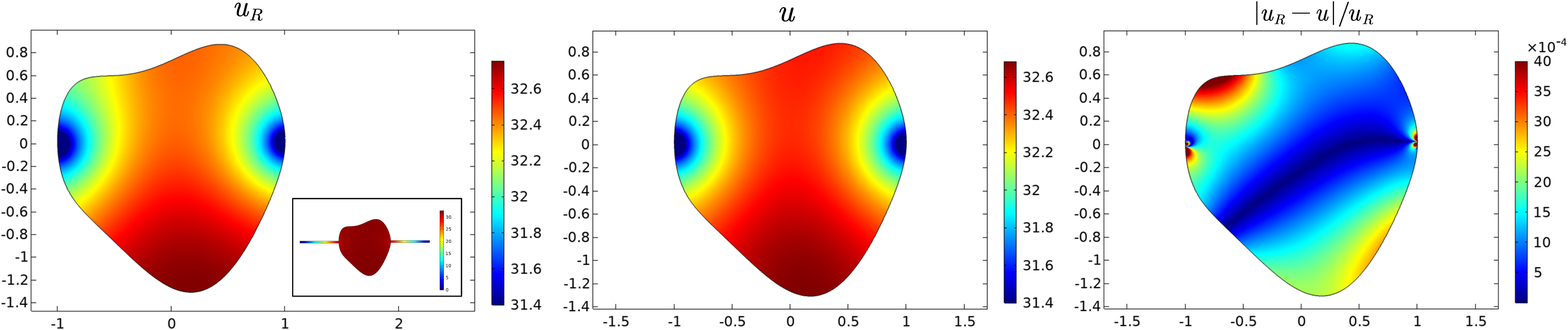}
\caption{General shaped domain with two necks. Neck length $L=1.5$, neck width $2\epsilon=0.08$.}
\end{subfigure}
\begin{subfigure}{1\textwidth}
\centering
\includegraphics[width=150mm]{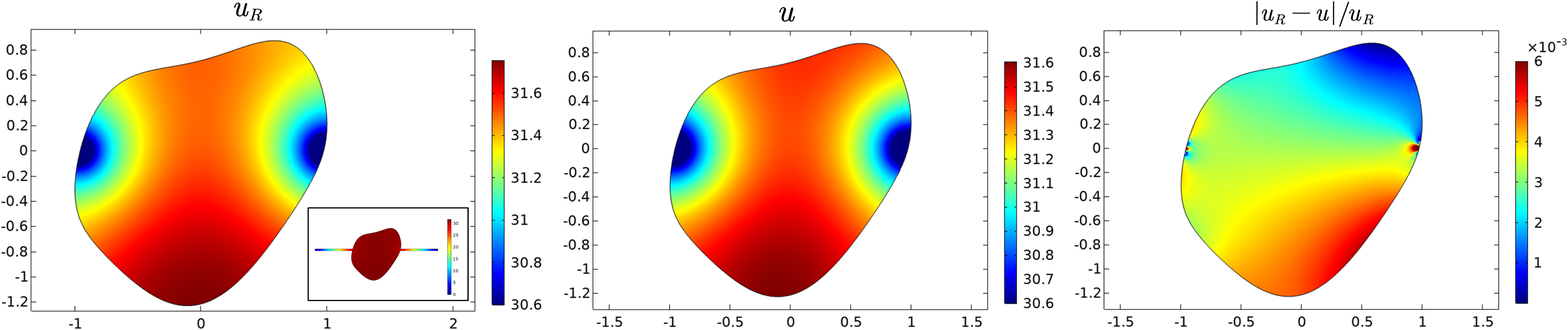}
\caption{General shaped domain with two necks. Neck length $L=1.5$, neck width $2\epsilon=0.08$.}
\end{subfigure}
\caption{The numerical solution $u_R$ (first column) , the asymptotic solution $u$ (second column) and  the relative error $|u_R-u|/u_R$  (third column) for two arbitrary shaped domain with two narrow necks.}
\label{fig:ire12}
\end{figure}

\begin{figure}[!ht]
  \centering
  \includegraphics[width=150mm]{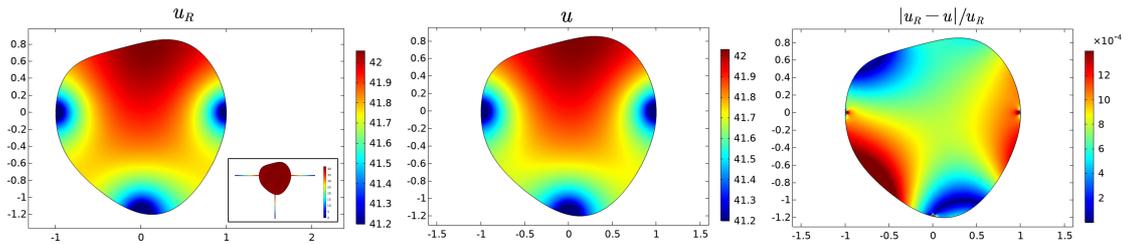}
  \caption{General shaped domain with three necks. Neck length $L=1.5$, neck width $2\epsilon=0.04$.}
  \label{fig:ire_3n}
\end{figure}

Finally, we conduct numerical experiments for general shaped domain $\Omega_h \in C^2(\mathbb{R}^2)$. The numerical solution $u_R$, the asymptotic solution $u$, and their relative error are shown in Figure \ref{fig:ire12} where two necks have the same neck length $L_1=L_2=1.5$ and width $2\epsilon_1=2\epsilon_2=0.08$. The parametrization of the head domain in Figure \ref{fig:ire12}(a) is: $x=\cos(t+\pi/3), y=\sin(t+\pi /3)-1/6\sin 2t+1/12\cos4t-1/12, t\in [0,2\pi)$. The parametrization of the head domain in Figure \ref{fig:ire12}(b) is: $x=\cos(t+\pi/4), y=\sin(t+\pi /3)-1/10\sin2t+1/15\cos4t-1/12, t\in [0,2\pi)$. Figure \ref{fig:ire_3n} shows the result that three necks are connected to the head domain, with the same length $L=1.5$ and the same width $2\epsilon=0.04$. The parametrization of the head domain in Figure \ref{fig:ire_3n} is: $x=\cos(t+\pi/3), y=\sin(t+\pi /3)-1/10\sin2t+1/27\cos4t-1/12, t\in [0,2\pi)$. The relative error $|u_R-u|/u_R$ shown in the third column, is very small which demonstrates that our asymptotic formula $u$ provides a good approximation to the MFPT for general shaped domain with several absorbing necks, which is the main result of our paper. 

\section{Conclusion}

In this study, we consider the narrow escape problem in a composite domain which consists of relatively big head of general shape and several absorbing thin necks. This is a follow up paper of \cite{Li}, where the Neumann-Robin model is proposed to solve the NEP in a relatively big head domain with only a single absorbing narrow neck. In this study, we derived three-term asymptotic expansion for the MFPT in terms of solving an equivalent Neumann-Robin model by layer potential techniques. The three-term asymptotic expansion reveals that the MFPT is determined by the lengths and radii of the necks as well as the nonlinear interaction between the necks. The first order term of the MFPT is proportional to the length $L$ and inversely proportional to the radius $\epsilon$, which is different from the well-known leading term $O(\ln \epsilon)$ for two dimensional NEP without necks. Due to the existence of narrow necks, it takes longer time for the particle to escape, which is natural. When $N$ well-separated narrow necks of the same length and the same radius are connected to the head domain, the first two terms of the asymptotic expansion is $1/N$ of that for a single neck gate derived in \cite{Li}. However the third term does not satisfy the $1/N$ relation due to the  interaction between necks. When two narrow necks are not well-separated, we investigate that the nonlinear interaction between two clustered necks affects not only the $O(1)$ term, but also the second term $O(\ln\epsilon)$, which has been explicitly reported in this paper. The three-term asymptotic expansion has been confirmed by the numerical results. Our asymptotic expansion with three leading order terms could approximate the MFPT up to error $O(\epsilon \ln^2\epsilon)$, which has not been reported previously. The study on solving the MFPT in a three dimensional domain connected with several thin tubular necks will be in a forthcoming paper.


\section*{Acknowledgement}
The authors would like to express their gratitude to the anonymous referees for their kind and useful comments. The work of Xiaofei Li was supported by NSF of China grant No. 11901523.


\end{document}